\documentstyle[psfig]{elsart}

\def\eg{{\it e.g.}}
\def\etal{{\it et al.}}

\def\gray{$\gamma$-ray}
\def\grays{$\gamma$-rays}

\def\mpc{\,{\rm Mpc}}

\def\fun#1#2{\lower3.6pt\vbox{\baselineskip0pt\lineskip.9pt
  \ialign{$\mathsurround=0pt#1\hfil##\hfil$\crcr#2\crcr\sim\crcr}}}
\newcommand{\beq}{\begin{equation}}
\newcommand{\eeq}{\end{equation}}
\begin{document}


\begin{frontmatter}

\title{Can Gamma Ray Bursts Produce the Observed Cosmic Rays Above $10^{20}$
eV?}
   
\author{F. W. Stecker}

\address{Laboratory for High Energy Astrophysics, Code 661,
NASA/Goddard Space Flight Center, Greenbelt, MD 20771, USA.}

\vspace{0.8cm}




\begin{abstract}

It has been suggested that cosmological \gray\ bursts (GRBs) can produce the
observed flux and spectrum of cosmic rays at the highest energies. However,
recent observations indicate that the redshift distribution of GRBs most
likely follows that of the star formation rate in the universe, a rate which
was much higher at redshifts 1.5-2 than it is today. Thus, most GRBs are at
high redshifts. As a consequence, any cosmic rays emitted by these GRBs
at energies above $\sim 2-3 \times 10^{19}$ eV would be strongly attenuated
by interactions with the 3 K background radiation. If one assumes rough
equality between the energy released by GRBs in $\sim 10^{-2}$ to $\sim 1$ 
MeV photons and that released in $10^{20}$ eV cosmic rays, then less than 
10\% of the cosmic rays observed above $10^{20}$ eV can be accounted for by 
GRBs.

\vskip 18pt

\end{abstract}

\begin{keyword}
gamma-ray bursts; cosmic rays; theory
\end{keyword}
\end{frontmatter}

\section{Introduction}

It has been suggested that cosmological \gray\
bursts can produce the observed flux of cosmic rays at the highest energies
\cite{wax95}, \cite{vie95}.
The arguments as stated in Ref. \cite{wax95} rest on four assumptions: 
(A) the highest energy cosmic rays are 
extragalactic, (B) cosmic rays can be accelerated to these energies in 
\gray\ burst fireballs, (C) the energy emitted by the bursts in ultrahigh
energy cosmic rays is roughly equal to the electromagnetic energy emitted
by the bursts (primarily in hard X-rays and soft \grays), and (D) the 
bursts have comoving density distribution which is independent of redshift,
{\it i.e.} there is no cosmological evolution.

While neither accepting or addressing assumptions (B) and (C), in this paper, 
I will argue that assumption (D) has become implausible when
one considers the recent redshift information obtained by locating the
afterglow radiation from the bursts in host galaxies with measured redshifts.
These studies place almost all \gray\ bursts (GRBs) with redshift assignments 
at moderate or high redshifts. Host galaxy studies imply that the GRB redshift 
distribution should follow the strong redshift dependence of the star
formation rate in galaxies (see section 3 below). A further 
implication is that the spatial density 
of \gray\ bursts at low redshifts would be too low to produce the observed flux
of cosmic rays above $10^{20}$ eV, since those cosmic rays can only reach us 
unattenuated in energy from distances of $\sim 100$ Mpc or less 
\cite{st68}, \cite{st99}, corresponding to redshifts 
$z \le \sim 10^{-2}$.

\section{The Energetics Argument for Non-evolving GRBs}

If one assumes that GRBs have a redshift independent co-moving distribution,
the energetics argument \cite{wax95},\cite{wax95b} 
can be summarized quite 
succinctly. If one takes the observed rate of GRBs and averages it out over
the volume of the observable universe, one finds an average rate per unit
volume of $r_{GRB} \simeq 1.5\times 10^{-8}$ \mpc$^{-3}$yr$^{-1}$ (taking
the Hubble constant in units of 100 km s$^{-1}$Mpc$^{-1}$, $h_{0} = 0.7$. 
If one  then takes an average total energy release per burst of 
$\sim4 \times 10^{52}$ erg in \grays\ (from the no evolution model of
Schmidt \cite{sch99}) and equates this to the energy released in 
ultrahigh energy cosmic 
rays, one finds a cosmic-ray energy input rate into intergalactic space of 
$\sim6\times 10^{44}$ erg \mpc$^{-3}$yr$^{-1}$.

Taking the differential cosmic ray spectrum given by Takeda, \etal\ 
\cite{tak98}, 
which fits a $E^{-2.78}$ power law for energies above $10^{19}$ eV, one finds
a cosmic ray energy flux between $10^{20}$ eV and $3\times 10^{20}$eV of 
$\Phi_{20} = 1.7$ erg m$^{-2}$sr$^{-1}$yr$^{-1}$. Using the similar power-law
spectrum given by Bird \etal\ \cite{bi93}, one finds an identical result. 
Taking a mean propagation distance of $ L \le 100$
Mpc for cosmic rays with energies above $10^{20}$ eV \cite{st68}, one then
finds that the required cosmic ray energy generation rate per unit volume
required to explain the flux of  cosmic rays in the 1-3 $\times10^{20}$ eV 
energy range is ($4\pi\Phi_{20})$/$L$ $\ge 2.1\times 10^{44}$ erg 
Mpc$^{-3}$yr$^{-1}$. 

The numbers given at the ends of the last two paragraphs are interestingly
similar. Thus, if as previously postulated, 
{\it e.g.}, \cite{wax95},\cite{wax95b}, a substantial fraction 
of the total GRB energy 
is released in ultrahigh energy cosmic rays as in \grays, 
GRBs can account for the observed particles above 
$10^{20}$ eV. As we will see, however, this argument is invalidated if one 
takes account of the redshift distribution of GRBs.

\section{The Redshift Distribution of GRBs and its Implications}  

The advent of the {\it BeppoSAX} X-ray telescope and the discovery of GRB 
X-ray \cite{co97}, optical \cite{ga97}, and radio 
\cite{fr97} afterglows and the subsequent identification 
of host galaxies has led
to the determination of the redshifts of some 11 GRBs from 1997 to date. 
Of these, 10 are at moderate to high redshifts and the remaining 
one, GRB980425,  has been identified with a nearby unusual Type Ic supernova, SN 1998bw \cite{ga98} with an energy
release ($\sim 5\times 10^{47}$ erg) which is orders of magnitude smaller
than the typical cosmological GRB. (In fact, it is not completely established
whether the supernova was indeed the source of the GRB, as another fading
X-ray source was a possible contender \cite{pi99}).
The GRB with the highest identified redshift to date, GRB971214, 
lies at a redshift of 3.42 \cite{ku98}. 

The positions of the bursts within the host galaxies
and their apparent association with significant column densities of hydrogen
and evidence of associated dust extinction \cite{ku98}, \cite{re98} 
has led to their  association with regions of active star 
formation.
Analyses of the colors of various host galaxies of GRBs has indicated that 
these galaxies are sites of active star formation \cite{ku98},
\cite{co98}, \cite{fr99} and this conclusion 
is strengthened by morphology studies and the detection of [OII] and 
Ly$\alpha$ emission lines in several host galaxies \cite{ku98}, \cite{me97},
\cite{bl98}. 

The association of GRBs with active star formation, together with the known 
strong redshift evolution of the star formation rate (\eg, \cite{ma98}) 
has led to theoretical examinations
testing whether a uniform comoving density redshift distribution or one which 
follows the star formation rate fits the GRB data best \cite{to97}- 
\cite{mm98}.
Mao \& Mo \cite{mm98} give a  discussion of the 
nature of the host galaxies of GRBs and argue for strong redshift evolution
of GRBs. The general conclusions of Mao \& Mo \cite{mm98} regarding the 
redshift distribution of GRBs are further supported in the most recent work 
\cite{sch99}, \cite{ko99},\cite{bu99}.

\section{GRB Redshift Evolution Leads to a Strong Energetics Problem}

Mao \& Mo \cite{mm98} find that their best fit model corresponds 
to a GRB redshift
distribution following the star formation rate which would have a 
{\it present rate} ($z \simeq 0$) of $\simeq 
1.7\times 10^{-10} h_{0}^3$ \mpc$^{-3}$yr$^{-1}$ and a mean energy release of 
$\sim 10^{52}h_{0}^{-2}$ erg per burst in the 50 to 300 keV band. 
Using more recent data,
Schmidt \cite{sch99} has given an analysis of the luminosities and space 
densities 
of GRBs. His analysis also points to a strong evolution in redshift,
similar to that of the star formation rate. He finds a present local GRB rate 
per unit volume of $\simeq 1.8\times 10^{-10}$ \mpc$^{-3}$yr$^{-1}$ with 
$h_{0}$ taken to be 0.7. Schmidt also finds a characteristic
total energy release per burst of $1.2 \times 10^{53}$ erg over the energy 
range from 10 to 1000 keV.  I will adopt Schmidt's more recent results 
\cite{sch99} for my discussion in this paper. 
The corresponding energy release rate per unit volume 
would then be $\sim 2\times 10^{43}$erg Mpc$^{-3}$yr$^{-1}$.
{\it This is an order of magnitude below the rate needed to explain the
ultrahigh energy cosmic rays, as indicated in Section 2 above.}
Therefore, even if we make the assumption of a rough equality
between the typical energy released by a GRB in $\gamma$-rays and that
released in ultrahigh energy cosmic-rays \cite{wax95},\cite{wax95b}, we still
fall significantly short of the energy input rate needed to explain the 
cosmic ray observations.

Another way of stating this result is that for GRBs to be the source of
the observed cosmic rays above $10^{20}$ eV, they would have to put at least
an order of magnitude more energy into $\sim 10^{14}$ MeV protons than into 
$\sim$ MeV photons. This would increase the required {\it total} GRB energy
to $ \ge 10^{54}$ erg and require GRBs to release at least 90\% of their energy
in the form of ultrahigh energy protons.

\section{Other Considerations}

There are other considerations which support the thesis presented
here that the GRBs are unlikely to produce the observed ultrahigh 
energy cosmic rays. 
Beaming is not a way out. While it is true that if GRBs are beamed into a
solid angle $\Omega$, we only see ($\Omega/4\pi$) of them, the energy release
per burst would also be lower by the same factor of $\Omega/4\pi$ and the total
energy release rate per unit volume is unchanged. Also, if the evolving 
redshift distribution scenario for GRBs
is correct, there will not be large numbers of faint GRBs nearby; the faintest
GRBs seen will corrrespond to GRBs which are at the highest redshifts.
(Even if the redshift distribution of bursts were more uniform than the star 
formation rate assumed here, this would imply that the average energy release 
per burst would be lower in order to fit the observed flux distribution, 
since there would be more nearby sources.)

Could Type I supernovae produce the observed ultrahigh energy cosmic rays? 
Let us assume that SN 1998bw is the source of GRB980425 and that some 
fraction of Type I SN are \gray\ bursters 
with a typical energy of $\sim 5 \times 10^{47}$ erg 
and a peak flux of $\sim 3 \times 10^{-7}$ erg cm$^{-2}$s$^{-1}$
in the {\it BATSE} range (as per GRB980425). Given its threshold flux, 
{\it BATSE} would be able to detect such sources distributed uniformly 
at a maximum rate of $\sim 60$ yr$^{-1}$ (taking the upper limit of 6\%
of the total burst rate given in Ref. \cite{ki98})
out to a distance of $\sim 53$ Mpc. The 
corresponding energy release rate would then be $\sim 3 \times 10^{49}$ erg
in a volume of $(4\pi)/3\times(53)^3$ Mpc$^3$ or $\sim 5\times10^{43}$ erg 
Mpc$^{-3}$yr$^{-1}$. This is only a factor of $\sim 4$ lower than the 
required rate (see section 2).
However, this is an upper limit, given the
statistical arguments against this hypothesis associating GRBs with Type I 
supernovae \cite{ki98} and \cite{gr99}. In addition, one may note that while 
a typical GRB has a ``high energy'' photon spectral index of 2.1, GRB980425
had a spectral index above $\sim 150$ keV of $\sim 4$, calling into question
whether such a source could produce ultrahigh energy cosmic rays.

\section{The Spectrum of Ultrahigh Energy Cosmic Rays}

Finally, I wish to comment on the spectrum of cosmic rays seen above 
$10^{20}$ eV. Waxman \cite{wax95b}
has argued that the present cosmic ray data may be still statistically
consistent with a uniform GRB distibution in redshift, even though no 
cosmological cutoff is seen corresponding to the so-called GZK effect
\cite{gr66}, \cite{za66}, \cite{st68}.
The GZK effect should
manifest itself in a steepening of the cosmic ray spectrum above an energy of
$\sim 7\times 10^{19}$ eV (\eg , \cite{st89}). If, as argued here however, the
GRBs are cosmic-ray sources overwhelmingly at moderate to high redshifts, 
the GZK effect comes in at lower energies (by a factor of $(1+z)^2$) and 
the attenuation will
be much more severe since the GZK process involves cosmic ray energy loss 
from photopion production off the 3K cosmic background radiation (which would
actually have a temperature of $3[1+z]$K) and the photon (target) density of 
this background would be higher by a factor of $(1+z)^3$). One thus expects to 
see a dramatic reduction in the observed flux above $\sim10^{19}$eV and 
{\it no} observable $10^{20}$ eV cosmic rays except those coming from
redshifts, $z \ll 1$ (see, {\it e.g.}, \cite{be88},\cite{yo93}). 
This is in strong contradiction to the observations
(\cite{ha94},\cite{bi95},\cite{tak98}.
This drastic conflict between the observed spectrum and that predicted for the
redshift distribution of GRBs will be presented in detail in a subsequent 
paper \cite{st00}.

\section{Conclusion}

Given all of the above considerations, it would appear that there is no
compelling reason to believe that GRBs can produce the observed flux of
ultrahigh energy cosmic rays. Indeed, given the knowledge obtained from
recent observations of GRBs, there appear to be many problems with this 
hypothesis, making it highly questionable. 

\section{Acknowledgments}

I would like to thank Robert Preece for a helpful discussion of the latest
GRB data. I would also like to thank Ralph Wijers for his helpful comments.


\newpage


\begin{thebibliography}{Costander}

\bibitem{wax95} Waxman, E. 1995, Phys. Rev. Letters 75, 386

\bibitem{vie95} Vietri, M. 1995, ApJ 453, 883

\bibitem{st68} Stecker, F.W. 1968, Phys. Rev. Letters 21, 1016

\bibitem{st99} Stecker, F.W. \& Salamon, M.H. 1999, ApJ 512, 521

\bibitem{wax95b} Waxman, E. 1995, ApJ 452, L1

\bibitem{sch99} Schmidt, M. 1999, ApJ 523, L117

\bibitem{tak98} Takeda, M. \etal\ 1998, Phys. Rev. Letters 81, 1163

\bibitem{bi93} Bird, D.C. \etal\ 1993, Phys. Rev. Letters 71, 3401

\bibitem{co97} Costa, E. \etal\ 1997, IAU Circ. No. 6576

\bibitem{ga97} Galama, T.J. \etal\ 1997, IAU Circ. No. 6584

\bibitem{fr97} Frail, D.A. \etal\ 1997, Nature 389, 261

\bibitem{ga98} Galama, T.J. \etal\ 1998, Nature 395, 670

\bibitem{pi99} Pian, E. \etal\ 1999, e-print astro-ph/9910235, submitted to ApJ

\bibitem{ku98} Kulkarni, S.R., \etal\ 1998, Nature 393, 35

\bibitem{re98} Reichert 1998, ApJ 495, L99

\bibitem{co98} Costander, F.J. \& Lamb, D.Q. 1998, in {\it Gamma Ray Bursts},
   ed. C.A. Meegan, R.D. Preece \& T.M. Koshut (New York: AIP) 520

\bibitem{fr99} Fruchter, A.S., \etal\ 1999, e-print astro-ph/9807295,ApJ, 
in press.

\bibitem{me97} Metzger, M., \etal\ 1997, Nature 387, 878

\bibitem{bl98} Bloom, J. \etal\ 1998, ApJ 507, L25

\bibitem{ma98} Madau, P., Pozzetti,L. \& Dickenson,M. 1998, ApJ 498, 106 

\bibitem{to97} Totani, T. 1997, ApJ 486, L71

\bibitem{to98} Totani, T. 1998, e-print astro-ph/9805263

\bibitem{wi98} Wijers, R. \etal\ 1998, MNRAS 294, L13

\bibitem{kr98} Krumholz, M., Thorsett, S.E. \& Harrison, F.A. 1998, 
ApJ 506, L81

\bibitem{mm98} Mao, S. \& Mo, H.J. 1998, A \& A 339, L1

\bibitem{ko99} Kommers, J.M. \etal\ 1999, ApJ, in press (e-print astro-ph/9809300)

\bibitem{bu99} Bulik, T. 1999, Gamma-Ray Bursts: The First Three Minutes,A.I.P. Conf. Ser. Vol. 190, 219

\bibitem{ki98} Kippen, R.M., \etal\ 1998, ApJ Letters 506, L27

\bibitem{gr99} Graziani, C.Lamb, D.Q., \& Marion, G.H. 1999, Astron \&
Ap Suppl Ser 138, 469 (see also e-print astro-ph/9810374)

\bibitem{gr66} Greisen 1966, Phys. Rev. Letters 16, 748

\bibitem{za66} Zatsepin, G.T. \& Kuzmin, V.A. 1966, JETP Letters 4, 78

\bibitem{st89} Stecker, F.W. 1989, Nature 342, 401

\bibitem{be88} Berezinsky, V.S. \& Grigor'eva, S.I. 1988, Astron. \& Ap. 199, 
1

\bibitem{yo93} Yoshida, S. \& Teshima, M. 1993, Prog. Theo. Phys. (Japan) 89, 
833

\bibitem{ha94} Hayashida, N. \etal\ 1994, Phys. Rev. Letters 73, 3491

\bibitem{bi95} Bird, D.C. \etal\ 1995, ApJ 441, 144

\bibitem{st00} Stecker, F.W. \& Scully, S. 2000, in preparation



\end{thebibliography}
\end{document}